\documentclass[aps,prb,twocolumn,nofootinbib,superscriptaddress]{revtex4-2}
\usepackage[dvipdfmx]{graphicx}
\usepackage{dcolumn}
\usepackage{times}
\usepackage{color}
\usepackage{bm}
\usepackage{braket}
\usepackage{physics}
\usepackage[version=4]{mhchem}
\usepackage{amsmath, amssymb}
\usepackage[colorlinks=true, citecolor=blue]{hyperref}

\usepackage{multirow}
\newcommand{\nn}{\nonumber}

\hypersetup{
setpagesize=false,
 bookmarksnumbered=true,%
 bookmarksopen=true,%
 colorlinks=true,%
 linkcolor=black,
 citecolor=blue,
}

\begin{document}
\preprint{APS/123-QED}
\title{Two-dimensional flat band on the (011) surface of UTe$_2$:\\
Implication for STM measurements with a superconducting tip}
\author{Jushin Tei}
\email{tei@blade.mp.es.osaka-u.ac.jp}
\affiliation{Department of Materials Engineering Science, The University of Osaka, Toyonaka 560-8531, Japan}
\author{Takeshi Mizushima}
\affiliation{Department of Materials Engineering Science, The University of Osaka, Toyonaka 560-8531, Japan}
\author{Satoshi Fujimoto}
\affiliation{Department of Materials Engineering Science, The University of Osaka, Toyonaka 560-8531, Japan}
\affiliation{Center for Quantum Information and Quantum Biology, The University of Osaka, Toyonaka 560-8531, Japan}
\affiliation{Center for Spintronics Research Network, Graduate School of Engineering Science, The University of Osaka, Toyonaka 560-8531, Japan}
\affiliation{Division of Spintronics Research Network, Institute for Open and Transdisciplinary Research Initiatives, The University of Osaka, Toyonaka 560-8531, Japan}
\date{\today}

\begin{abstract}
        Scanning tunneling microscopy (STM) measurements have been extensively performed on the easily cleavable $(011)$ surface of UTe$_2$, using both normal-metal and superconducting tips.
        Motivated by these experiments, we theoretically investigate the topological surface states on the $(011)$ surface of UTe$_2$.
        We find that a two-dimensional nearly flat band emerges in the $B_{3u}$ state, giving rise to a pronounced zero-energy peak in the surface density of states.
        This flat band is supported by two key mechanisms: (i)~nontrivial Berry phases defined at multiple momenta give rise to low-energy in-gap states, and (ii)~weak spin conservation allows the gap function to acquire phase winding.
        Furthermore, to investigate the relation between the zero-bias peak observed in recent STM experiments with a superconducting tip and the topological surface states, we calculate the nonequilibrium dc tunneling current in a junction between an $s$-wave superconductor and the $(011)$ surface of UTe$_2$.
        Our results provide crucial insights into the superconducting pairing symmetry realized in UTe$_2$.
\end{abstract}

\maketitle

\section{Introduction}
The uranium-based superconductor UTe$_2$ has emerged as one of the most enigmatic superconductors discovered in recent years, widely considered a candidate for spin-triplet $p$-wave pairing due to its strikingly unconventional properties~\cite{ran2019Nearly,aoki2019Unconventional,aoki2022Unconventional,ran2019Extreme,rousal2023field,zheyu2024enhanced,braithwaite2019Multiple,aoki2020Multiple,knebel2020Anisotropy,aoki2021FieldInduced}.
A central topic is the nature of its pairing symmetry, as it governs the structure of low-energy excitations and, crucially, dictates the realization of topological superconductivity in UTe$_2$. 
Various experimental studies~\cite{fujibayashi2022superconducting,matsumura2023Large,aishwarya2023magnetic,lee2023anisotropic,suetsugu2024fully,hayes2024robust,li2025observation,theuss2024single}
and theoretical works~\cite{ishizuka2021periodic,shishidou2021topological,kreisei2022spin,yu2023theory,hakuno2024magnetism,haruna2024possible,tei2024pairing,crepieux2025quasiparticle,chiristiansen2025nodal} have been conducted to elucidate the pairing symmetry.
Despite these extensive efforts, the issue remains unresolved.

On the surface of a topological superconductor, zero-energy states---known as Andreev bound states (ABS)---emerge as a hallmark of its topological nature~\cite{hu1994midgap,kashiwaya2000tunnelling,sato2009topological,sato2010topological,sato2011topology,alicea2012new,sato2016majorana}, and the emergence of ABS directly influences the surface density of states (DOS).
This topological character is fundamentally linked to the odd-parity symmetry of the gap function, which satisfies the Andreev condition $\Delta({\bm k}) = -\Delta(-{\bm k})$, where $\Delta({\bm k})$ is the superconducting gap at momentum ${\bm k}$.
Consequently, the appearance of ABS is intimately tied to the superconducting pairing symmetry and surface orientation.
Scanning tunneling microscopy (STM), which directly probes the surface DOS, therefore offers a powerful means to identify the pairing symmetry in unconventional superconductors.

Recently, STM experiments were performed on the naturally cleavable $(011)$ surface of UTe$_2$,
using both a normal-metal tip~\cite{jiao2020chiral,gu2023detection,aishwarya2023magnetic,yoon2024probing,sharma2025observation,yin2025yin} and a superconducting tip~\cite{gu2023detection,gu2025pair,wang2025imaging}.
Remarkably, a sharp zero-bias peak (ZBP) was observed in the differential conductance $(dI/dV)$ spectrum, comparable in magnitude to the total coherence peaks~\cite{gu2025pair,wang2025imaging}.
In a junction between a spin-singlet $s$-wave superconductor and the spin-triplet superconductor UTe$_2$, single-quasiparticle tunneling is suppressed due to the fully gapped excitation spectrum of the $s$-wave electrode, and the Josephson current {is highly suppressed} owing to the symmetry mismatch in spin space.
This suggests that the observed ZBP most likely originates from Andreev reflection mediated by ABS of UTe$_2$.
However, two major questions remain.
First, if the ZBP is attributed to the surface DOS of the $(011)$ surface, it is unclear why such a large number of zero-energy ABS exist to produce a pronounced zero-energy peak in the surface DOS.
Second, since there has been no microscopic calculation of the nonequilibrium dc tunneling current between a topological superconductor and an $s$-wave superconductor, interpreting the $dI/dV$ spectra
has remained an open issue, especially in the context of Andreev spectroscopy.
Regarding the first question, it should be emphasized that typical topological surface states in three-dimensional (3D) spin-triplet superconductors, which is called Fermi arcs, cannot lead to a sharp ZBP in $dI/dV$ spectra~\cite{yang2014dirac,sch15}.
In a superconducting state with pairwise point nodes, the Fermi arc is a zero-energy flat band forming the segment that connects the point nodes on its surface Brillouin zone (BZ).
However, such a 1D flat band does not result in a peak structure at zero-energy but yields a constant surface state density that is independent of the energy.
Notable exceptions include line-nodal superconductors~\cite{kashiwaya1995origin,wei1998directional,iguchi2000angle,kashiwaya2000tunnelling} and superfluids~\cite{dmitiev2015polar,zhelev2016observation,kamppinen2023topological}, as well as odd-parity superconducting states realized in carrier-doped topological insulators~\cite{hao2011surface,hsieh2012majorana,yamakage2012theory,sasaki2012odd,sasaki15}.
In the former, the combination of bulk line nodes and the formation of ABS leads to 2D flat bands, resulting in a sharp zero-energy peak in the surface DOS.
In the latter, the dispersion of the ABS is intertwined with that of the surface Dirac cone inherited from the parent topological insulator, forming a smooth spectral connection between them and thereby enhancing the zero-energy surface DOS.
Since UTe$_2$ is believed to be either a point nodal or a fully gapped superconductor, and its normal state is topologically trivial, these mechanisms are not directly applicable.

In this work, we theoretically investigate the ABS on the $(011)$ surface of UTe$_2$ for all possible irreducible representations (IR) of the pairing symmetry---$A_u$, $B_{1u}$, $B_{2u}$, and $B_{3u}$.
The classification of surface states in topological superconductors protected by crystalline symmetries has been extensively studied, and a complete classification has already been mostly established~\cite{chiu2014classification,shiozaki2014topology,shiozaki2016topology}.
To exhibit a zero-energy peak in the surface DOS, however, a mechanism beyond the conventional arguments based solely on topological invariants is required.
Remarkably, we find that a 2D nearly flat band emerges only in the $B_{3u}$ state.
The sufficient amount of zero-energy states extended over the 2D surface BZ results in a pronounced zero-energy peak in the surface DOS. 
We note that the surface states in the other pairing states are dispersive and, no sharp peak structure appears in their surface DOS.
The emergence of the 2D flat band in the $B_{3u}$ state originates from two distinct mechanisms intrinsic to its pairing state:
(1) nontrivial Berry phases defined at multiple high-symmetry momentum points not only guarantee the presence of zero-energy states, but also give rise to low-energy in-gap states that smoothly connect them; and
(2) weak spin conservation intrinsic to the $B_{3u}$ state, which allows the gap function to acquire nontrivial phase winding, thereby generating additional ABS protected by a 1D winding number.

We then explore the relation between the observed ZBP and the emergence of the 2D nearly flat band.
To this end, we numerically compute the tunneling current between UTe$_2$ and an $s$-wave superconducting tip using the tunnel-Hamiltonian method.
In this calculation, we fully incorporate all orders of tunneling processes, including resonant transmission and multiple reflections.
Furthermore, for qualitative insight, we derive an analytical expression for the fourth-order Andreev current.
In the low-bias regime, where the derived expression is valid and relevant to STM measurements, the $dI/dV$ characteristics are shown to be directly determined by the convolution of the surface DOS of UTe$_2$.
Finally, by comparing our theoretical results with the STM experiments using a superconducting tip, we discuss the superconducting pairing symmetry realized in UTe$_2$.

The organization of this paper is as follows.
In Sec.~\ref{Sec2}, we introduce the model Hamiltonian for the superconductor of UTe$_2$.
In Sec.~\ref{Sec3}, we present numerical results for the quasiparticle energy spectra and the surface DOS of the $(011)$ surface for all IR, together with a topological argument that accounts for the emergence of ABS.
In Sec.~\ref{Sec4}, we numerically calculate the tunneling current between an $s$-wave superconductor and the $(011)$ surface of UTe$_2$.
The detailed algorithm used for calculating tunneling current is provided in Appendix~\ref{App:C}.

\begin{figure}[t]
        \centering
        \includegraphics[width=\linewidth]{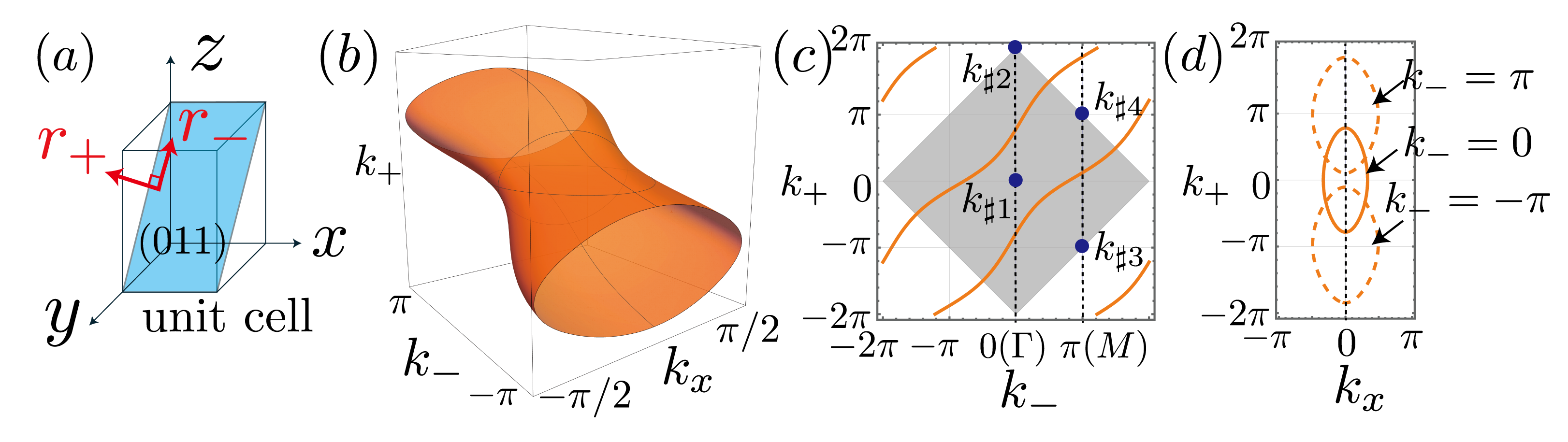}
        \caption{
        (a)~A naturally cleavable $(011)$ plane. 
        (b)~Cylindrical Fermi surface elongated along the $k_z$-axis,
        plotted in the rotated coordinate system $(k_x, k_+, k_-)$,
        where the $k_yk_z$-plane is rotated by $\pi/4$ around the $k_x$-axis.
        (c)~Plot of the $k_x = 0$ plane in the first BZ (shaded area),
        with the Fermi surface shown as orange curves.
        $k_{\sharp i}$~($i = 1 \sim 4$) denotes the time-reversal-invariant momenta (TRIMs).
        Nontrivial Berry phases can be defined along 1D paths in $k_+ \in [-2\pi, 2\pi]$,
        centered at $\Gamma = (k_x = 0, k_m = 0)$ and $M = (k_x = 0, k_m = \pi)$.
        Each path connects a pair of TRIMs lying along the $k_+$ direction.
        (d)~Fermi surface plots at $k_m = 0$ (solid curves) and $k_m = \pm\pi$ (dashed curves).
        }
        \label{fig:model}
\end{figure}

\section{Model of superconductor UTe$_2$\label{Sec2}}
We begin by describing a minimal model Hamiltonian for UTe$_2$.
This compound crystallizes in a body-centered orthorhombic lattice with the point group $D_{2h}$.
The geometry of the Fermi surface plays a crucial role in realizing topological superconductivity.
Recent de Haas-van Alphen experiments have observed Fermi surfaces, 
which consist of two cylindrical sheets---one electron-like and one hole-like---extending along the 
$k_z$-direction~\cite{aoki2022First,eaton2024quasi}. 
To capture the essential features, we consider a tight-binding model on a simple orthogonal lattice with a Kramers-degenerate spin degree of freedom.
The normal-state Hamiltonian is given by 
$H_{\rm N}(\bm{k}) = 2t_1 \cos k_x + 2t_2 \cos k_y + 2t_3 \cos k_z - \mu$.
To reproduce the cylindrical Fermi surface shown in Fig.~\ref{fig:model}(b), we set the parameters as
$t_1 = -1.0,~$$t_2 = -1.0,$~$t_3 = 0.25$, and $\mu = -2.5$.
Although this minimal model is adopted for clarity, in Appendix~\ref{App:B}, 
we have verified that our main results remain robust against moderate variations in the parameters,
as well as in more realistic models that incorporate orbital degrees of freedom, detailed Fermi surface topology, and staggered Rashba spin-orbit coupling (SOC) arising from local inversion symmetry breaking.
As demonstrated in the following, the cylindrical geometry of the Fermi surface plays a particularly crucial role in the formation of a 2D nearly flat band.

The gap function for spin-triplet Cooper pairs is described by the $d$-vector: 
$\hat{\Delta}(\bm{k}) = \bm{d}({\bm k})\cdot\boldsymbol{\sigma}i\sigma_y$.
The point group $D_{2h}$ allows four odd-parity IR:
\begin{align}
        &\bm{d}_{A_u}(\bm{k}) = 
               ( C_x\sin k_x, C_y \sin k_y, C_z \sin k_z
        ), \\
        &\bm{d}_{B_{1u}}(\bm{k}) = 
                ( C_x\sin k_y, C_y \sin k_x, C_z \sin k_x \sin k_y \sin k_z
        ), \\
        &\bm{d}_{B_{2u}}(\bm{k}) = 
                ( C_x\sin k_z, C_y \sin k_x \sin k_y \sin k_z, C_z \sin k_x
        ), \\
        &\bm{d}_{B_{3u}}(\bm{k}) = 
                (C_x\sin k_x\sin k_y \sin k_z,  C_y \sin k_z,  C_z \sin k_y).
\end{align}
The $A_u$ state corresponds to a fully gapped superconducting state.
The $B_{1u}$ state is also fully gapped, due to the absence of a Fermi surface along the $k_z$ axis.
In contrast, the $B_{2u}$ and $B_{3u}$ states are Dirac superconductors with point nodes along the $k_x$ and $k_y$ axes, respectively.

In this study, we focus on the $(011)$ surface as illustrated in Fig.~\ref{fig:model}(a).
To facilitate the analysis,
we introduce a rotated coordinate system $(x,r_+,r_-)$, where the $yz$-plane is rotated by $\pi/4$ around the $x$-axis.
The corresponding rotated momenta are defined as $k_+ = k_y + k_z$ and $k_- = -k_y + k_z$.
Since the $(011)$ surface is perpendicular to the $r_+$ direction;  
the momenta $k_x$ and $k_-$ are conserved.
Figures~\ref{fig:model}(b-d) show the BZ in the rotated coordinate frame.

\section{Topological surface state on (011) surface\label{Sec3}}

\begin{figure*}[t]
        \centering
        \includegraphics[width=\linewidth]{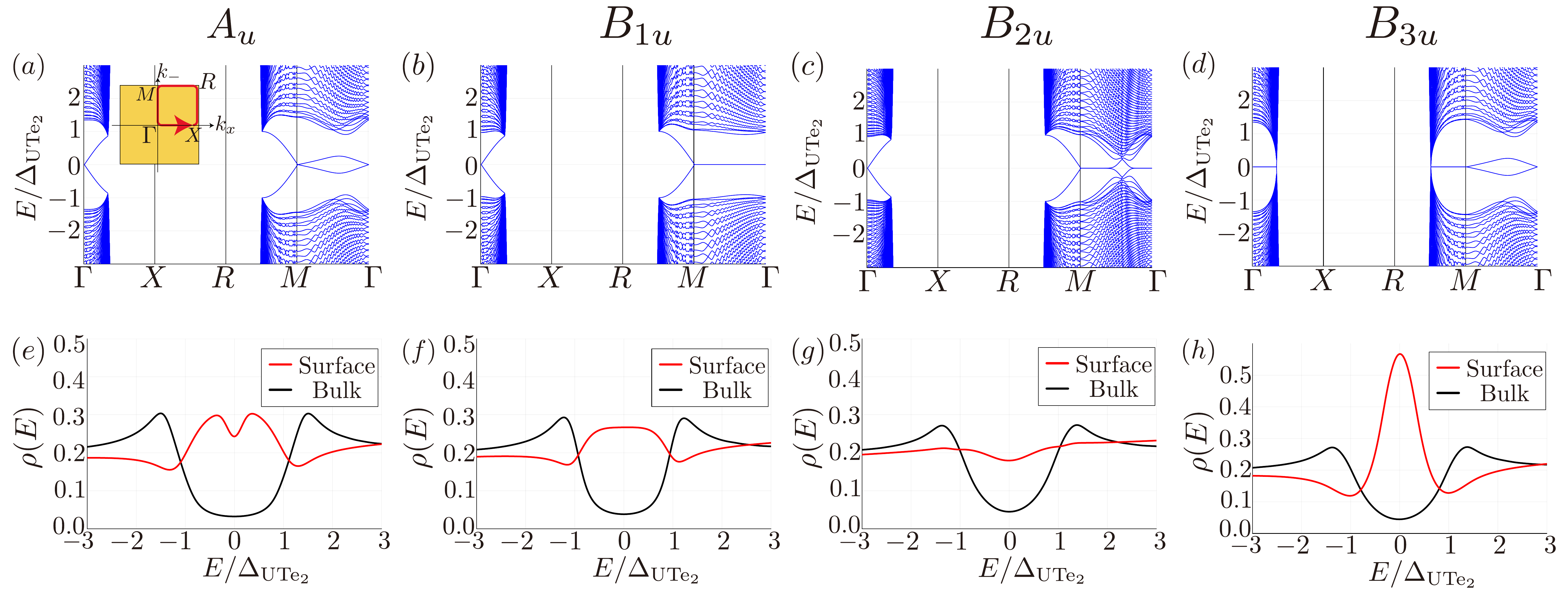}
        \caption{(a–d)~Quasiparticle energy bands for the (a)~$A_u$, (b)~$B_{1u}$, (c)~$B_{2u}$, and (d)~$B_{3u}$ states.
        The superconducting gap $\Delta_{\mathrm{UTe}2}$ is set to $0.05$.
        The inset in panel~(a) shows the momentum path in the surface BZ depicted in the main panels.
        For the $A_u$ state, zero-energy states protected by nontrivial Berry phases appear at the $M$ and $\Gamma$ points, and low-energy in-gap states smoothly connecting them emerge along the $M$–$\Gamma$ line.
        For the $B{1u}$ and $B_{2u}$ states, Fermi arcs protected by the one-dimensional winding number associated with mirror reflection symmetry appear.
        For the $B_{3u}$ state, in addition to the in-gap states along the $M$–$\Gamma$ line, Fermi arcs protected by the one-dimensional winding number associated with spin conservation emerge along the $\Gamma$–$X$ and $R$–$M$ lines.
        (e–h)~DOS for the $(011)$ surface (red) and the bulk (black) for the (e)~$A_u$, (f)~$B_{1u}$, (g)~$B_{2u}$, and (h)~$B_{3u}$ states.
        For the $B_{3u}$ state, as a result of the spread of nearly zero-energy states across the 2D surface BZ, the surface DOS exhibits a pronounced zero-energy peak.}
        \label{fig:allIR}
\end{figure*}
We diagonalize the Bogoliubov-de Gennes (BdG) Hamiltonian that consists of $H_{\rm N}({\bm k})$ and $\hat{\Delta}({\bm k})$ in a slab system with the opened $(011)$ surfaces.
Figures~\ref{fig:allIR}(a-d) present the quasiparticle energy spectra along the symmetric path in the surface BZ, 
$\Gamma(0,0)-X(\pi,0)-R(\pi,\pi)-M(0,\pi)-\Gamma(0,0)$, 
where $(k_x, k_m)$ represents the momenta in the surface BZ for all IR.
We also compute the surface DOS using the recursive Green's function method~\cite{umerski1997closed,ohashi2024anisotropic}, as shown in Figs.~\ref{fig:allIR}(e-h).
In this calculation, the smearing factor $\delta$, which reflects the quasiparticle lifetime,
is set to $\delta = 0.2\Delta_{\text{UTe}_2}$, where $\Delta_{\text{UTe}_2}=0.05$.
Note that $\Delta_{\text{UTe}_2}$ serves as a prefactor [$C_x, C_y$, $C_z$] in the $d$-vector, and does not represent the actual superconducting gap magnitude.
In the following, we explain the origin of the ABS for each IR.

\subsection{The $A_u$ state}
The non-chiral $A_u$ superconducting state belongs to class DIII in the Altland-Zirnbauer classification~\cite{schnyder2008classification}.
Since this state is fully gapped, a 3D winding number can be defined as the associated topological invariant.
However, it remains trivial as long as the Fermi surface retains a quasi-2D cylindrical shape.
Thus, weak or crystalline topological invariants are expected to play a crucial role in the emergence of ABS.

Figures~\ref{fig:allIR}(a) and \ref{fig:allIR}(e) show the quasiparticle energy bands and the corresponding surface DOS, respectively.
In this calculation, the parameters are set to $C_x = 0.05,~C_y = 0.05$, and $C_z = 0.05$.
Zero-energy states appear at the high-symmetry momentum points $\Gamma$ and $M$.
At these points, the Berry phase can be defined as a $\mathbb{Z}_2$ topological invariant.
Specifically, we consider two one-dimensional loops along $k_+\in[-2\pi,2\pi]$ 
passing through the $\Gamma$ and $M$ points, respectively, as illustrated by the dashed lines in Fig.~\ref{fig:model}(c)~\cite{tei2023possible}.
The Berry phase along each loop is determined by the parity of the number of Fermi surfaces that intersect the loop between two time-reversal-invariant momenta, denoted $k_{\sharp i}$. 
When the number of intersecting Fermi surfaces is odd (even), the Berry phase is nontrivial (trivial)~\cite{sato2009topological,sato2010topological}.
As a consequence, topologically protected zero-energy states appear at both the $\Gamma$ and $M$ points.

A further noteworthy feature is the presence of in-gap states that are isolated from the bulk spectra along the $M-\Gamma$ line.
The zero-energy states protected at the $M$ and $\Gamma$ points do not merge directly into the bulk band; instead, they connect smoothly as in-gap states.
Such surface states lead to the DOS exhibiting a dome-like structure inside the superconducting gap.
The broad split peaks indicate the presence of the van Hove singularities along the $M-\Gamma$ direction.
Notably, this type of ABS is unique to the $(011)$ surface.
We emphasize that the formation of these isolated in-gap states requires the presence of zero-energy states protected by Berry phases at an even number of $k$-points.
This feature can also be related to the triviality of the mirror Chern number defined on the $M-\Gamma$ line:
in order for the surface states to remain isolated from the bulk bands, the absence of net charge pumping is required, which is equivalent to the mirror Chern number being zero~\cite{fu2006time,fu2007topological,thouless1983quantization}.
This condition naturally arises from the cylindrical Fermi surface.
However, the triviality of the Chern number is not a sufficient condition for the existence of the isolated in-gap states.
In this sense, the in-gap states are not topologically protected; rather, they are model-specific and strongly depend on the details of the pairing interaction.

\subsection{The $B_{1u}$ and $B_{2u}$ states}
The $B_{1u}$ state is a fully gapped superconducting state since the cylindrical Fermi surface is opened along the $k_z$ axis.
Figure~\ref{fig:allIR}(b) shows the quasiparticle energy bands,
and Fig.~\ref{fig:allIR}(f) displays the surface DOS for the $B_{1u}$ state.
In this calculation, the parameters are set to $C_x = 0.05,~C_y = 0.05$, and $C_z = 0.0$.
Unlike the $A_u$ state, a perfectly flat band appears along the $M-\Gamma$ line.
As a result, the energy dependence of the surface DOS remains nearly constant around zero energy.
In the symmetry class DIII, the only nontrivial topological invariant is the Berry phase at the high-symmetry $M$ and $\Gamma$ points.
Therefore, this flat band must originate from additional topological invariants associated with crystalline symmetries---in particular, the mirror symmetry $\mathcal{M}_{yz}$.
In the following, we describe the topological invariants associated with mirror symmetry~\cite{tei2023possible}.

The mirror operator $\mathcal{M}_{yz}$ acts on the normal-state Hamiltonian ${H}_{\rm N}$ as
\begin{eqnarray}
  \mathcal{M}_{yz}^{-1}H_{\rm N}(k_x,k_+,k_-)\mathcal{M}_{yz} = H_{\rm N}(-k_x,k_+,k_-), 
\end{eqnarray}
and on the gap function $\Delta$ as
\begin{eqnarray}
  \mathcal{M}_{yz}^{-1}\Delta(k_x,k_+,k_-)\mathcal{M}_{yz} = s\Delta(-k_x,k_+,k_-),
\end{eqnarray}
where $s = + 1$ for the $B_{1u}$ and $B_{2u}$ states, and
$s = -1$ for the $A_u$ and $B_{3u}$ states.
Depending on the value of $s$, the mirror operator acting on the Bogoliubov-de Gennes (BdG) Hamiltonian is given by
\begin{eqnarray}
  \tilde{\mathcal{M}}_{yz} = \begin{pmatrix}
    \mathcal{M}_{yz} & \\
    & s\mathcal{M}_{yz}^{\ast}
  \end{pmatrix}.
\end{eqnarray}
On the mirror plane ($k_x = 0$), this operator $\tilde{\mathcal{M}}_{yz}$ commutes with the BdG Hamiltonian.
We can then define a crystalline chiral operator that anti-commutes with the BdG Hamiltonian:
\begin{eqnarray}
  \Gamma_{\mathcal{M}_{yz}} = e^{i\phi}\tilde{\mathcal{M}}_{yz}\Theta C,
  \label{eq:gamma}
\end{eqnarray}
where $e^{i\phi}$ is a phase factor chosen such that $\Gamma_{\mathcal{M}_{yz}}^2 = 1$,
$\Theta$ is the time-reversal operator, 
and $C$ is the particle-hole exchange operator. 
The properties of the chiral operator defined in Eq.~\eqref{eq:gamma} depend on the value of $s$.
For $s = -1$, corresponding to the $A_u$ and $B_{3u}$ states,
the chiral operator anti-commutes with the time-reversal operator $\Theta$,
which is the same as the standard chiral operator in class DIII, $\Gamma = i\Theta C$.
In contrast, for $s=+1$, as in the $B_{1u}$ and $B_{2u}$ states,
the chiral operator commutes with the time-reversal operator---characteristic of class BDI.
In class BDI, a 1D winding number can be defined using the chiral operator as a topological invariant:
\begin{align}
  w_{\mathcal{M}_{yz}}(k_-) = -\frac{1}{4\pi i}\int dk_+~\text{tr}~[\Gamma_{\mathcal{M}_{yz}}\mathcal{H}_{\text{BdG}}\partial_{k_+}\mathcal{H}_{\text{BdG}}],
  \label{eq:winding}
\end{align}
where $\mathcal{H}_{\rm BdG}$ is the BdG Hamiltonian that is composed of $H_{\rm N}$ and $\hat{\Delta}({\bm k})$. Note that for the $A_u$ and $B_{3u}$ states,
this 1D winding number is not a valid topological invariant, 
as it always vanishes due to the anti-commutation between the chiral and time-reversal operators.
The winding number in Eq.~\eqref{eq:winding} can be efficiently calculated using Fermi surface formula. For both the $B_{1u}$ and $B_{2u}$ states, it simplifies to
\begin{eqnarray}
  w_{\mathcal{M}_{yz}}(k_-) = \sum_{E_{\rm N}(k_+)=0} \text{sgn}[d_x]\text{sgn}[\partial_{k_+}E_{\rm N}],
\end{eqnarray}
where $E_{\rm N}({\bm k})$ is the eigenvalue of the normal-state Hamiltonian ${H}_{\rm N}({\bm k})$.
This means that for a fixed $k_-$ on the $k_x = 0$ plane,
the winding number can be obtained by summing the products of the signs of the $d_x$ component and the Fermi velocity $\partial_{k+}E_{\rm N}$
at each Fermi crossing point along the 1D loop $k_+\in[-2\pi,2\pi]$.
The resulting winding number is $w=2$, reflecting the twofold degeneracy of the flat band.

As mentioned above, in the $B_{2u}$ state, the mirror symmetry $\mathcal{M}_{yz}$ also gives $s=+1$, allowing the 1D winging number to be defined as a topological invariant.
Therefore, the $B_{2u}$ state is qualitatively similar to the  $B_{1u}$ state, except for the presence of Dirac points of the bulk state.
Figure~\ref{fig:allIR}(c) shows the quasiparticle energy band,
and Fig.~\ref{fig:allIR}(g) displays the surface DOS for the $B_{2u}$ state.
In this calculation, we set $C_x = 0.05,~C_y = 0.0$, and $C_z = 0.05$.
Along the $M-\Gamma$ line, both the Dirac points and ABS protected by the 1D winding number appear.
The surface DOS exhibit a shallow dip structure, reflecting the coexistence of bulk Dirac bands and the 1D flat band.

\subsection{The $B_{3u}$ state}
\begin{figure}[t]
        \centering
        \includegraphics[width=\linewidth]{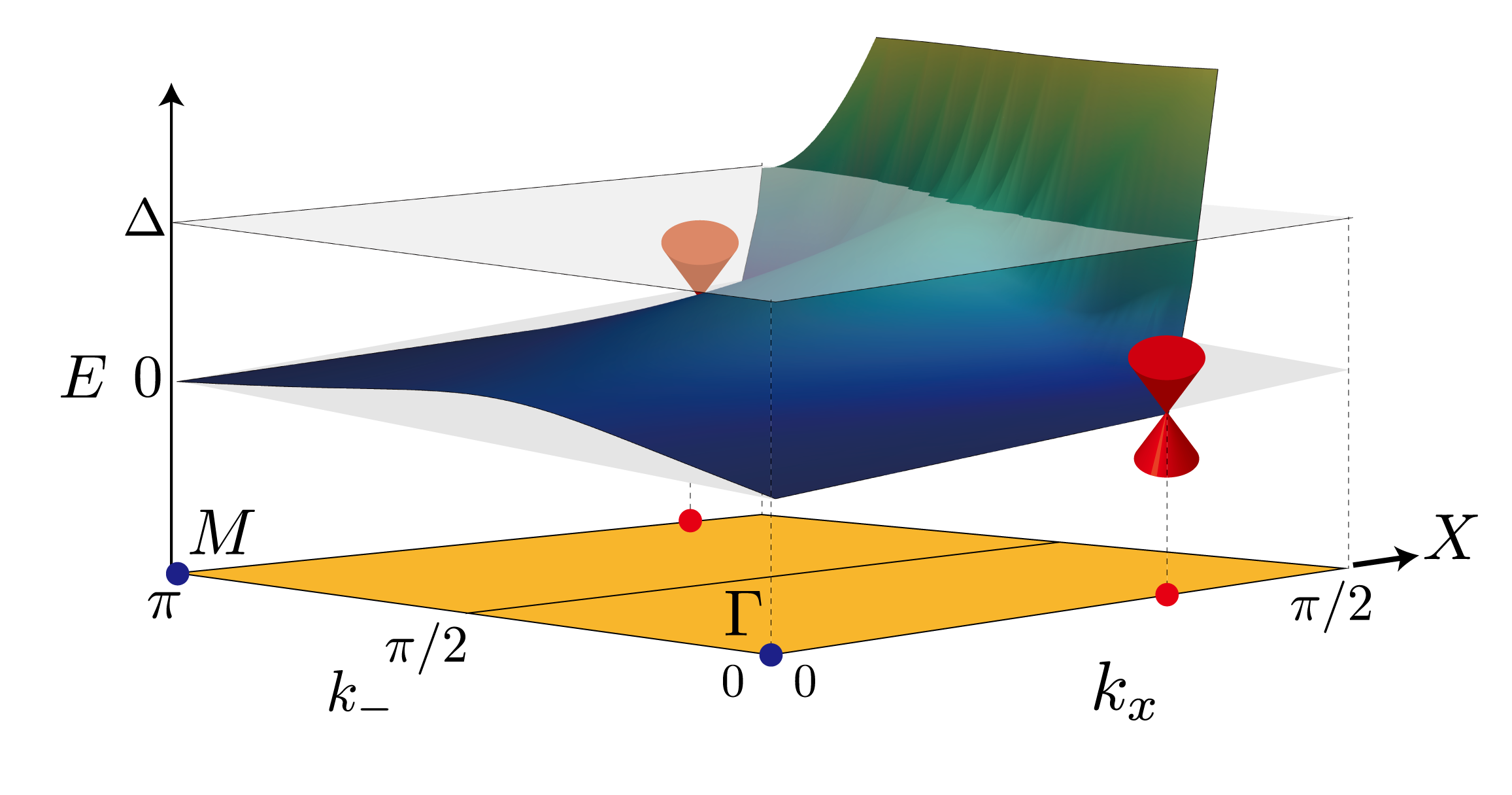}
        \caption{Quasiparticle energy bands on the $(011)$ plane for the superconducting $B_{3u}$ state. 
        A substantial number of ABS form a 2D nearly flat-band structure.
        The red cones indicate the point nodes of the bulk superconducting gap.}
         \label{fig:flatland}
\end{figure}
Figure~\ref{fig:allIR}(d) displays the quasiparticle energy bands,
and Fig.~\ref{fig:allIR}(h) shows the surface DOS for the $B_{3u}$ state.
In this calculation, we set $C_x = 0,~C_y = 0.05$, and $C_z=0.05$.
Remarkably, we observe the emergence of zero-energy flat bands along the $\Gamma-X$ and $R-M$ lines, 
as well as an almost flat band near zero energy along the $M-\Gamma$ line.
These low-energy ABS extend across the surface BZ as a 2D nearly flat band, as illustrated in Fig.~\ref{fig:flatland}.
As a result, surface DOS exhibits a pronounced zero-energy peak.
In the following, we discuss the topological origin of such a 2D flat band.

The nearly zero-energy ABS along the $M$–$\Gamma$ direction corresponds to the in-gap state found in the $A_u$ state.
Specifically, given the triviality of the Chern number, zero-energy states protected by Berry phases exist at the $M$ and $\Gamma$ points, and the in-gap state connects them smoothly.
The origin of another set of zero-energy ABS along the $\Gamma-X$ and $R-M$ lines lies in a weak spin conservation intrinsic to the $B_{3u}$ state.
When the $d_x$ component vanishes $(C_x = 0)$, the gap function can be diagonalized by a spin rotation, yielding $\hat{\Delta} = \text{diag}(d_z+id_y,~-d_z+id_y)$.
This form reveals spin conservation, where the spin quantization axis is along the $x$-direction.
If we consider only up-spin Cooper pairs, the gap function reduces to
\begin{eqnarray}
        \Delta_{\uparrow\uparrow} \propto e^{i\pi/4}\sin\frac{k_+}{2}\cos\frac{k_-}{2}
        - e^{-i\pi/4}\cos\frac{k_+}{2}\sin\frac{k_-}{2} .
\end{eqnarray}
Within the up-spin subspace, we can define a winding number $w(k_x,k_m)$ associated with the chiral symmetry along the 1D path $k_+\in[-2\pi,2\pi]$ as a function of $k_x$ and $k_m$.
This winding number can be evaluated using Fermi surface formula
\begin{eqnarray}
        w = \frac{1}{2}\sum_{E_N(k_+)=0}\text{sgn}[\Delta_{\uparrow\uparrow}]\text{sgn}[{\partial_{k_+}E_N}].
\end{eqnarray}
As shown in Fig.~\ref{fig:model}(d),
along the $\Gamma-X$ line ($k_m=0$), the Fermi surface is symmetric concerning $k_+$,
and the gap function $\Delta_{\uparrow\uparrow}\propto \sin(k_+/2)$ is an odd function of $k_+$.
Consequently, the winding number is $w=1$ whenever the 1D $k_+$ path crosses the Fermi surface.
Similarly, the winding number is $w=1$ along the $R-M$.

\begin{figure}[t]
        \centering
        \includegraphics[width=\linewidth]{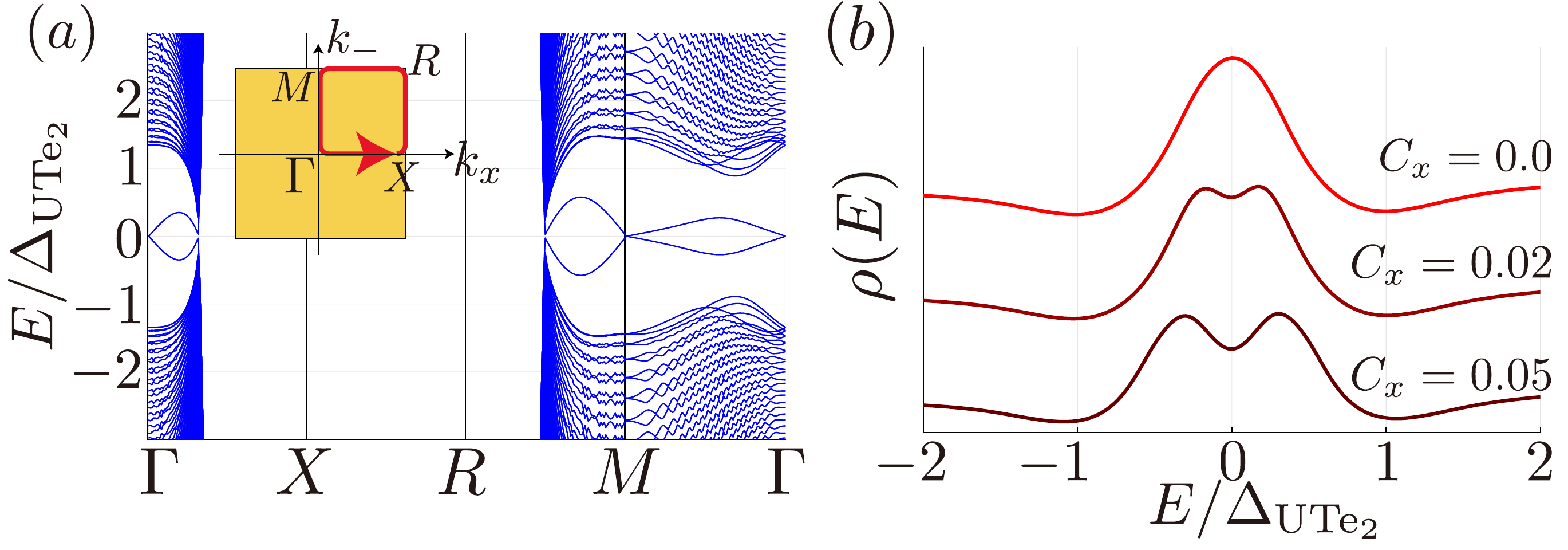}
        \caption{(a)~Quasiparticle energy bands for the $B_{3u}$ state with parameters $C_x = 0.05$, $C_y = 0.05$, and $C_z = 0.05$.
        $\Delta_{{\rm UTe}_2}$ is set to $0.05$.
        The $d_x$ component breaks spin conservation, resulting in a gap opening along the $\Gamma$--$X$ and $R$--$M$ lines
        (b)~Surface density of states as a function of $C_x$.}
        \label{fig:Cx}
\end{figure}
From the above discussion, the ABS along the $\Gamma-X$ and $R-M$ lines originate from accidental spin conservation and are therefore not protected by any symmetry.
In particular, the presence of a $d_x$ component violates spin conservation, leading to spin flip.
However, even in the presence of a finite $d_x$ component, in-gap states (not necessarily at zero energy) can still be robust, as we will discuss in the following.
Figure~\ref{fig:Cx}(a) shows the quasiparticle energy spectrum for the $B_{3u}$ state with pairing parameters set to $C_x = 0.05,~C_y = 0.05$, and $C_z = 0.05$.
The zero-energy flat bands that previously existed in the $\Gamma-X$ and $R-M$ lines in Fig.~\ref{fig:allIR}(d) are now gapped out.
As a result of spin flipping, the surface DOS exhibits a maximum at finite energy, corresponding to van Hove singularities of surface states,
as shown in Fig.~\ref{fig:Cx}(b).
Nevertheless, although the flat bands are lifted from zero energy,
a substantial number of ABS remain within the superconducting gap.
In this sense, weak spin conservation along the $x$-direction allows 2D nearly flat band to remain robust.
The extent to which these states stay close to zero energy depends on the specific model.
However, as long as the $d_x$ component remains sufficiently small---comparable to the energy broadening---a zero-energy peak in the surface DOS continues to be present.

We conclude that the 2D nearly flat band in the $B_{3u}$ state arises from the in-gap states induced by the cylindrical Fermi surface, together with the existence of a 1D winding number associated with spin conservation.
We emphasize that, as can be inferred from other IR, a 1D flat band originating from 1D winding number does not necessarily lead to a zero-energy peak in the surface DOS.
We comment on the role of SOC.
Although our starting model assumes Kramers spin degeneracy,
the effects of atomic SOC and staggered Rashba SOC---arising from local inversion symmetry breaking at the uranium sites---are effectively incorporated.
As a result, the 2D nearly flat band remains robust even in the presence of SOC.
In Appendix~\ref{App:B}, we explicitly demonstrate that the 2D nearly flat band persists in a two-orbital model that includes staggered Rashba SOC.

\section{Andreev Current Spectroscopy in UTe$_2$\label{Sec4}}
\begin{figure*}[t]
        \centering
        \includegraphics[width=\linewidth]{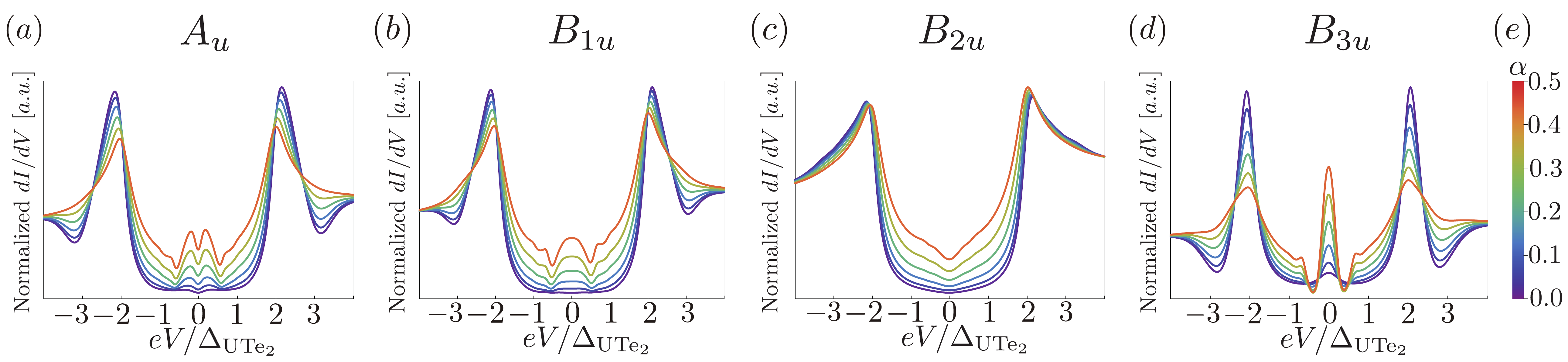}
        \caption{Normalized differential conductance $(dI/dV)$ of a superconducting junction between UTe$_2$ and an $s$-wave superconductor.
        The superconducting state in UTe$_2$ corresponds to the (a)~$A_u$, (b)~$B_{1u}$, (c)~$B_{2u}$, and (d)~$B_{3u}$ representations, respectively.
        The $s$-wave superconducting gap is set to $\Delta_s = 2\Delta_{\text{UTe}_2}$.
        (e)~Color bar shows the effective transparency, $\alpha \in [0,1]$.}
        \label{fig:dIdV}
\end{figure*}

To investigate recent STM experiments with an $s$-wave superconducting tip,
we analyze the $dI/dV$ characteristics of the tunneling current between an $s$-wave superconductor and each of the possible IR proposed for UTe$_2$.
The tunneling current consists of three main contributions.
The first is a single-particle tunneling, which vanishes at low bias $(eV<\Delta_{s})$
due to the absence of available quasiparticle states in the $s$-wave superconducting tip. 
The second contribution arises from Cooper pair tunneling
which is highly suppressed due to the spin-space orthogonality between spin-triplet and spin-singlet superconductors~\cite{ges86,mil88}.
Thus, the dominant contribution in the low-bias regime is Andreev tunneling.
In this process, electrons and holes in the ABS of the topological superconductor are converted into Cooper pairs in the $s$-wave superconductor and vice versa.  

We now compute the tunneling current starting from a microscopic Hamiltonian,
$\mathcal{H} = \mathcal{H}_{\text{UTe}_2} + \mathcal{H}_{s-\rm{wave}} + \mathcal{H}_{T}$, 
where the tunneling term is expressed as $\mathcal{H}_T = T\sum_{k,q}c^\dagger_{k}d_{q} + \text{h.c.}$.
Here, $T$ represents the tunneling matrix, and $c_{k}~(d_q)$ denotes the electron annihilation operator for UTe$_2$ (the $s$-wave superconducting tip).
Since tunneling occurs locally in real space, momentum is not conserved,
and the tunneling matrix is taken to be momentum-independent.
The tunneling current, defined as $I(V)= e\langle \dot{N} \rangle$, is evaluated by perturbation theory using Keldysh Green's function formalism~\cite{cuevas1996hamiltonian}.
The details are described in Appendix~\ref{App:C},
and the current is calculated using Eq.~\eqref{eqS:I}. 
We emphasize that Eq.~\eqref{eqS:I} gives an exact expression for the tunneling current, fully incorporating multiple reflections at the interface.
{As a consistency check, we explicitly confirm that our formulation reproduces the well-established dc tunneling conductance of conventional $s$-wave/$s$-wave superconducting junctions~\cite{cuevas1996hamiltonian}.}
As a convenient measure of the tunneling strength, we define the effective transparency, $\alpha \in [0,1]$, as  
\begin{eqnarray}
        \alpha = \frac{4\pi^2|T|^2{N_sN_{\rm UTe_2}}}{(1 + \pi^2|T|^2{N_sN_{\rm UTe_2}})^2},
\end{eqnarray}
where $N_s$ and $N_{\rm UTe_2}$ denote the DOS at the Fermi level in the normal state for the $s$-wave superconductor and UTe$_2$, respectively.
In STM experiments, $\alpha$ typically takes small values.
In the numerical calculations, the $s$-wave superconducting gap is set to $\Delta_s = 2\Delta_{\text{UTe}_2}$.

Figure~\ref{fig:dIdV} shows the normalized $dI/dV$ spectra from the weak to intermediate tunneling strength for all IR.
Remarkably, we observe an exceptionally sharp ZBP in the $B_{3u}$ state, with a magnitude comparable to that of the coherence peaks of the $s$-wave superconducting tip.
This $dI/dV$ spectrum is in excellent agreement with recent experimental observations~\cite{gu2025pair,wang2025imaging}.
To obtain qualitative insight into the low-bias regime at low transparency, we derive the expression for the Andreev tunneling current in the weak-coupling and low-bias limit:
\begin{align}
\label{eq:Andreev}
I(V) = \frac{4e}{\hbar}|T|^4\pi^3N_s^2\int^{2eV}_0 dE~\rho(E-2eV)\rho(E),
\end{align}
where $\rho(E)$ denotes the surface DOS in the superconductor UTe$_2$.
This originates from a fourth-order perturbation process in the tunnel Hamiltonian as reflected in $|T|^4$ dependence,
and it is independent of the $s$-wave superconducting gap $\Delta_s$.
Consequently, the appearance of a ZBP serves as a direct signature of the 2D flat band.
Importantly, such a pronounced ZBP in the $(011)$ surface is unique to the $B_{3u}$ state.
Thus, these results support the realization of the $B_{3u}$ pairing state in UTe$_2$.
However, it should be noted that no ZBP has been reported in STM experiments using normal-metal tips~\cite{gu2025pair,wang2025imaging},
even though the presence of the 2D flat band is expected to produce a similar signature regardless of the tip type.
This discrepancy underscores the need for a more careful and comprehensive examination of the origin of the ZBP and the underlying pairing symmetry in UTe$_2$.

We comment on the splitting of the ZBP of the conductance observed in Ref.~\cite{gu2025pair}.
The authors argue that, for non-chiral pairing, a superconducting phase difference between the $p$-wave and $s$-wave gaps breaks time-reversal symmetry, leading to a gap opening in the surface Majorana band and a splitting of the ZBP.
As can be seen from Eq.~\eqref{eq:Andreev}, however, the dc tunneling current is independent of the phase difference.
This conclusion is further supported by Eq.~\eqref{eqS:I}, which incorporates all tunneling processes up to arbitrary order [see Fig.~\ref{fig:ZBPsp}(b)].
These results indicate that further examination is required to clarify the origin of the ZBP splitting of the conductance.
{This difference arises from the following reason.
In Ref.~\cite{gu2025pair}, the conductance is calculated by using the Weidenmuller-type formula for the $S$-matrix. According to the standard scattering theory for junction systems, the energy levels that appear in the denominator of the Weidenmuller formula are those of a decoupled subsystem, i.e., the Majorana surface bands of isolated UTe$_2$, which is decoupled from the $s$-wave superconducting tip in our case. However, in Ref.~\cite{gu2025pair}, the energy levels
used in the $S$-matrix formula are those of the SS junction system with the phase difference between the $s$-wave gap and the $p$-wave gap, which results in the splitting of the ZBP of the conductance. 
In contrast, 
in Appendix~\ref{App:A}, we demonstrate that while the spectral density of the surface Majorana band depends on the superconducting phase difference, the dc tunneling current does not.}

\section{Conclusion and Remark}
Motivated by the observation of a pronounced ZBP in STM experiments with a superconducting tip, we have investigated ABS on the $(011)$ surface of UTe$_2$.
In the superconducting $B_{3u}$ state, a sufficient amount of nearly zero-energy states extends over the 2D surface BZ, resulting in a pronounced zero-energy peak in the surface DOS.
This 2D flat band originates not only from a 1D winding number associated with weak spin conservation but also from in-gap states connecting the zero-energy states protected by nontrivial Berry phases.
Furthermore, we have calculated the $dI/dV$ spectra of the dc tunneling current of a junction between an $s$-wave superconductor and UTe$_2$.
Only the $B_{3u}$ state, which realizes a 2D flat band on the $(011)$ surface, gives rise to a pronounced ZBP in the calculated $dI/dV$ spectra through the Andreev reflection process.
Therefore, the observation of a ZBP strongly supports the realization of the $B_{3u}$ state in UTe$_2$.
However, the absence of a ZBP in experiments using a normal-metal tip calls for further theoretical and experimental studies.

{We comment on the NMR experiments, which observed a large reduction of the Knight shift for a magnetic field along the $a$ axis, and its relation to the results of this paper~\cite{matsumura2023Large}.
Such a pronounced reduction may naively suggest a dominant $a$-axis component of the $d$-vector and thus appear inconsistent with a $B_{3u}$ state under spin conservation; however, the situation in UTe$_2$ is more subtle.
In particular, in the normal state just above the superconducting transition, the spin susceptibility along the $a$ axis is strongly enhanced due to ferromagnetic fluctuations with Ising anisotropy along the $a$ axis~\cite{ran2019Nearly}.
This strong enhancement may be substantially suppressed below $T_c$ because of the opening of the gap, leading to 
the large reduction of the Knight shift along the $a$ axis, irrespective of the direction of the $d$-vector.
Consequently, the $B_{3u}$ state is not necessarily incompatible with the NMR experiments.}

We also comment on the multiband effects that are neglected in the present model.
Since the observed ZBP requires a sufficient amount of zero-energy states, such as a 2D zero-energy flat band, it is unlikely that the multiband effect is the most essential factor in the formation of the ZBP.
Nevertheless, multiband effects may alter the surface states in UTe$_2$, especially at finite energies. 
A comprehensive discussion of the surface DOS in multiband superconductivity is beyond the scope of the present work and will be addressed in future studies.

Although our discussion has focused on UTe$_2$, the mechanism of the 2D flat band is broadly applicable.
In particular, the geometry of the Fermi surface gives rise to nontrivial Berry phases at multiple momenta, and the associated zero-energy states connect smoothly to form in-gap states.
A natural direction for future work is to generalize these results to other Fermi surface geometries and topological symmetry classes.
Furthermore, strong correlation effects in flat bands on topological superconducting surfaces present a particularly intriguing avenue for further exploration.
The surface of a topological superconductor is not a purely 2D system; rather, it constitutes an anomalous surface---one that inherits the Bogoliubov quasiparticles.  
Interaction-induced gapping without symmetry breaking could potentially give rise to new classes of topologically ordered phases, distinct from those realized in purely 2D electron systems~\cite{PhysRevX.5.041013}.

\begin{acknowledgments}  
The authors are grateful to X. Liu, T. Hanaguri, Y. Tanaka, K. Shiozaki, T. Matsushita, and R. Ohashi for fruitful discussions. 
J.T. is grateful to Q. Gu for valuable discussions from an experimental point of view.
J.T. is supported by a Japan Society for the Promotion of Science (JSPS) Fellowship for Young Scientists.
This work was supported by JSPS KAKENHI (Grant No.~JP23K20828, No.~JP23K22492, No.~JP24KJ1621, No.~JP25H00599, No.~JP25H00609, No.~JP25K07227, and No.~JP25K22011) and a Grant-in-Aid for Transformative Research Areas (A) ``Correlation Design Science'' (Grant No.~JP25H01250) from JSPS of Japan..
\end{acknowledgments}

\appendix

\section{Gap opening in the zero-energy flat band by a phase difference}
\label{App:A}
\begin{figure*}[t]
  \centering
  \includegraphics[width=\linewidth]{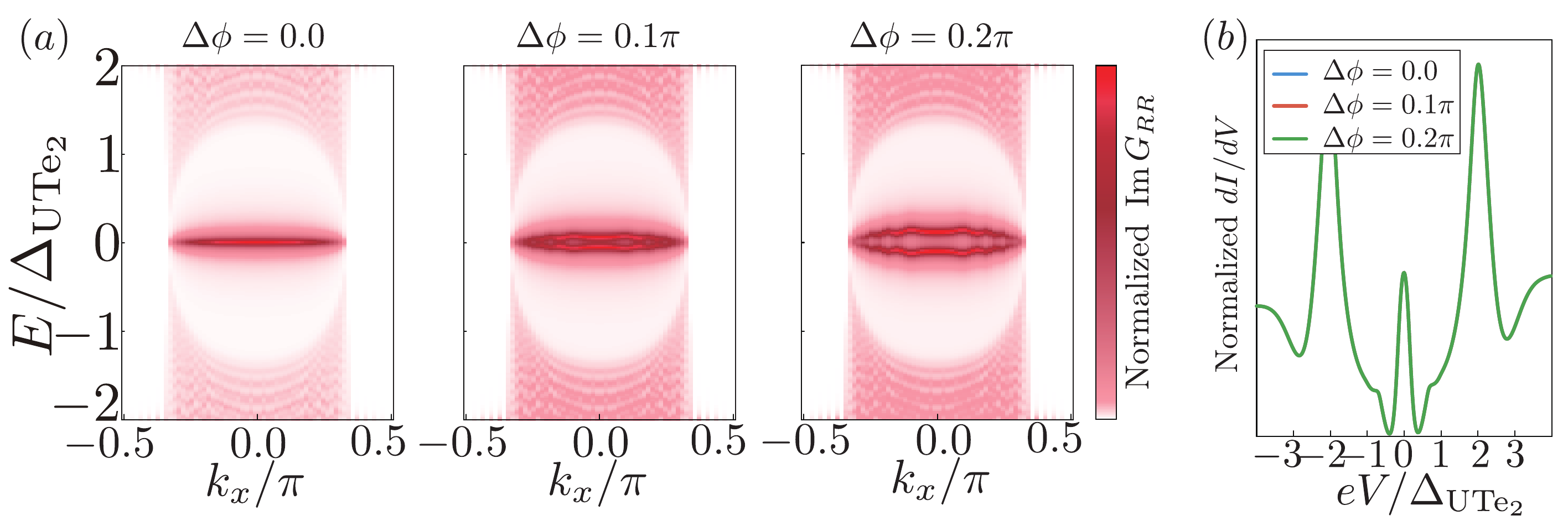}
  \caption{(a)~Momentum-resolved spectral density for the $B_{3u}$ state obtained from $G_{RR}$, which incorporates tunneling effects at the interface, as a function of superconducting phase difference $\Delta\phi$.
  The phase difference induces a gap opening in the Fermi arc along the $X$--$\Gamma$--$X$ line $(k_m=0)$. 
  (b)~Normalized $dI/dV$ of the dc tunneling current for $\Delta\phi = 0,~0.1\pi$, and $0.2\pi$; all curves coincide, showing no dependence on the superconducting phase difference.
  }
  \label{fig:ZBPsp}
\end{figure*}

Reference~\cite{gu2025pair} explains that the breaking of time-reversal symmetry induced by a superconducting phase difference, $\Delta\phi=\pi/2$, originating from the Josephson coupling, opens a gap in the surface Majorana band, leading to a splitting of the ZBP in the $dI/dV$ spectrum.
{We note that although in the absence of SOC, the second-order Josephson coupling between an $s$-wave and a $p$-wave superconductor vanishes, higher-order Josephson coupling terms can still exist. These higher-order terms favor a stable relative phase difference of $\pi/2$ in the $s$-wave/$p$-wave junction.}
On the other hand, our calculation of the dc tunneling current is independent of $\Delta\phi$.
At first glance, these results may appear contradictory; however, 
we demonstrate below that while the surface Majorana band depends on the phase difference, the dc tunneling current does not.

We here consider a planar junction between the $(011)$ surface of UTe$_2$ and an $s$-wave superconductor, referred to as the SIP model~\cite{gu2025pair}.
Note that in the main text, we have assumed a point-contact geometry for the STM junction.
The total Hamiltonian is given by $\mathcal{H} = \mathcal{H}_{L} + \mathcal{H}_{R} + \mathcal{H}_{T}$.
The right electrode corresponds to UTe$_2$, while the left electrode is the $s$-wave superconductor.
The tunneling Hamiltonian depends on the superconducting phase difference $\Delta\phi$ as
\begin{align}
    \mathcal{H}_{T} = |T|e^{i\Delta\phi/2}\sum_{k}c_{k}^{\dagger}d_{k} + {\rm h.c.},
\end{align} 
which represents momentum-conserving tunneling.
Here, $c_{k}~(d_k)$ and $c^{\dagger}_{k}~(d^{\dagger}_k)$ denote the electron annihilation and creation operators for the left (right) electrode, respectively.
Using perturbation theory with respect to $\mathcal{H}_T$, we calculate  the full Green's function $\check{G}_{RR}$, which incorporates all orders of tunneling processes.
From Eq.~\eqref{eq:GDyson}, we obtain
\begin{align}
    \check{G}_{RR} = \check{G}_{RR}^{(0)} + \check{G}_{RR}^{(0)}\check{T}^{\dagger}\check{G}_{LL}^{(0)}\check{T}\check{G}_{RR}
\end{align}
where $\check{G}_{RR}^{(0)}$ corresponds to the Green's function of $\mathcal{H}_R$ for UTe$_2$, reproducing surface DOS illustrated in Figs.~\ref{fig:allIR}(e-f),
and $\check{G}_{LL}^{(0)}$ denotes the Green's function of $\mathcal{H}_L$ for the $s$-wave superconductor.
Here, the symbol $\check{M}$ indicates that a matrix $M$ is expressed in the rotated Keldysh basis, which consists of retarded, advanced, and Keldysh components.

Figure~\ref{fig:ZBPsp}(a) shows the momentum-resolved spectral density incorporating tunneling effects along the $X$--$\Gamma$--$X$ line $(k_m=0)$ for the $B_{3u}$ state,
which is obtained from the full Green's function, $\Im G_{RR}(\bm{k},E)$, as a function of the phase difference.
When there is no phase difference, a Fermi arc protected by a 1D winding number appears.
However, when the phase difference is finite, the Fermi arc is lifted from zero energy due to the breaking of time-reversal symmetry. 
Furthermore, we calculate the dc tunneling current for a point-contact junction geometry as a function of phase difference, using the same procedure as in the main text.
As shown in Fig.~\ref{fig:ZBPsp}(b), the dc tunneling current is completely independent of the phase difference.
It should be noted that this difference does not originate from the junction geometry.
The tunneling current is expressed by $G_{LR}$, as shown in Eq.~\eqref{eqS:I}.
The total current can be decomposed into two contributions: a phase-dependent component (e.g., the Josephson current) and a phase-independent component corresponding to the dc tunneling current.
$G_{RR}$ is determined self-consistently through the Dyson equation in relation to $G_{LR}$, thereby taking into account both phase-dependent and phase-independent tunneling processes.
Therefore, the gap opening of the surface Majorana band arises as a consequence of the Josephson coupling, and the dc tunneling current is considered to be independent of it.

\section{Tunneling Hamiltonian approach\label{App:C}}
\subsection{General expression for the tunneling current by Keldysh Green's function}
In this section, we provide a brief derivation of the tunneling current in Josephson junctions
using the tunneling Hamiltonian approach~\cite{cuevas1996hamiltonian}.
The total Hamiltonian of the system consists of three terms:
\begin{eqnarray}
  &&\mathcal{H} = \mathcal{H}_{R} + \mathcal{H}_{L} + \mathcal{H}_{T}, \\
  &&\mathcal{H}_{T} = \sum_{k,q}T_{kq}d^{\dagger}_{k}c_q + T^{*}_{kq}c^{\dagger}_qd_k,
\end{eqnarray}
where $\mathcal{H}_{L(R)}$ denotes the Hamiltonian of the left (right) system, 
constructed from the electron annihilation operators $c_q$ ($d_k$) and creation operators $c^{\dagger}_q$ ($d^{\dagger}_k$).
The term $\mathcal{H}_{T}$ describes the tunneling between the two systems,
where $T_{kq}$ is the tunneling matrix element that reflects the geometric structure of the junction.
When an STM tip makes contact at a point, the tunnel matrix can be taken as momentum-independent, i.e., $T_{kq}=T$.
Although spin degrees of freedom are not explicitly included in the tunneling Hamiltonian,
we assume spin-conserving tunneling processes in the absence of SOC and magnetic impurities.

The voltage bias between the two systems can be incorporated as a shift in the chemical potential.
Taking the chemical potential of the right system as a reference, the Hamiltonian of the left system under a finite voltage bias $V$ is given by
\begin{eqnarray}
  \mathcal{H}_L(V) = \mathcal{H}_{L}(V = 0) + eVN_L,
\end{eqnarray}
where $N_L$ is the total number operator in the left system.
In the Heisenberg picture, the operators of the left system transform as
\begin{eqnarray}
  &&\tilde{c}_q(t) = e^{-\frac{i}{\hbar}eVt}c_{q}(t), \\
  &&\tilde{c}^{\dagger}_q(t) = e^{\frac{i}{\hbar}eVt}c^{\dagger}_{q}(t), 
\end{eqnarray}
where $c_q(t)$ and $c_q^{\dagger}(t)$ are the annihilation and creation operators in the Heisenberg picture without the voltage bias.

The tunneling current is defined as the time derivative of the electron number in the left system:
\begin{eqnarray}
  I(V,t) = e\langle \dot{N}_L(t)\rangle,
\end{eqnarray}
where 
\begin{eqnarray}
  \dot{N}_L &=& \frac{i}{\hbar}[\mathcal{H},N_{L}] = \frac{i}{\hbar}[\mathcal{H}_T,N_L] \nonumber \\
  &=& \frac{i}{\hbar} \sum_{k,q} \left(T_{kq}d^{\dagger}_{k}c_q - T^*_{kq}c^{\dagger}_qd_k\right).
\end{eqnarray}
Here, we used the fact that $N_L$ commutes with both $\mathcal{H}_{L}$ and $\mathcal{H}_R$.
With straightforward manipulation, the tunneling current can be expressed in terms of the lesser Green's function across the junction as
\begin{eqnarray}
  \label{eqS:I}
  I(V,t) = -2e \Re \sum_{k,q} \text{tr}~T(k,q) G^<_{LR}(q,t;k,t'),
\end{eqnarray}
where the trace is taken over spin and sublattice degrees of freedom.
We now introduce the time-ordered Green's function for the coupled junction system:
\begin{align}
    \hat{G}_{LR} = \begin{pmatrix}
        G^{11}_{LR} & G^{<}_{LR} \\
        G^{>}_{LR} & G^{22}_{LR}
    \end{pmatrix},
\end{align}
and
\begin{eqnarray}
  \label{eq:G11}
  &&G^{11}_{LR}(q,t;k,t') = -\frac{i}{\hbar}\langle T_{\text{c}}~ \tilde{c}_{q}(t)d^{\dagger}_k(t') \rangle, \\
  &&G^<_{LR}(q,t;k,t') = \frac{i}{\hbar}\langle  d^{\dagger}_k(t')\tilde{c}_{q}(t) \rangle,  \\
  &&G^{>}_{LR}(q,t;k,t') = -\frac{i}{\hbar}\langle \tilde{c}_{q}(t)d^{\dagger}_k(t') \rangle, \\
  \label{eq:G22}
  &&G^{22}_{LR}(q,t;k,t') = -\frac{i}{\hbar}\langle \bar{T}_{\text{c}}~ \tilde{c}_{q}(t)d^{\dagger}_k(t') \rangle, 
\end{eqnarray}
where $T_{\text{c}}~(\bar{T}_{\text{c}})$ denotes the contour (anti-contour) time-ordering operator in the Keldysh formalism.

\subsection{Perturbation theory for the Green's function of the junction}
To determine the Green's function of the junction, we treat the tunneling Hamiltonian as a perturbation.
For this purpose, it is convenient to introduce the rotated Keldysh basis for perturbation theory:
\begin{align}
  \check{G} \equiv \check{L}{\tau_3}\hat{G}\check{L}^\dagger = \begin{pmatrix}
    G^R & G^K \\
    0 & G^A
  \end{pmatrix},~~~
  \check{L} = \frac{1}{\sqrt{2}}({\tau_0} - i{\tau}_2),
\end{align}
where $ \boldsymbol{\tau}$ is the Pauli matrix acting of Keldysh space.
$G^R$, $G^A$, and $G^K$ denote retarded, advanced, and Keldysh components of Green's function, respectively:
\begin{widetext}
\begin{eqnarray}
  &&G^R(1,2) = G^{11}(1,2) - G^{<}(1,2) = G^{21}(1,2) - G^{22}(1,2) = -\frac{i}{\hbar}\langle \{c(1), c^\dagger(2) \} \rangle \theta(t_1-t_2), \\
  &&G^A(1,2) = G^{11}(1,2) - G^{>}(1,2) = G^{12}(1,2) - G^{22}(1,2) = \frac{i}{\hbar}\langle \{c(1), c^\dagger(2) \} \rangle \theta(t_2-t_1), \\
  &&G^K(1,2) = G^{<}(1,2) + G^{>}(1,2) = G^{11}(1,2) + G^{22}(1,2) = -\frac{i}{\hbar}\langle [c(1),c^\dagger(2)] \rangle,
\end{eqnarray}
where we used the shorthand notation $(1) = (q,t)$ and $(2) = (k,t')$.
In thermal equilibrium at inverse temperature $\beta$, the Keldysh component takes the form
\begin{eqnarray}
  G^{K}(q,\omega) = (G^{R}(q,\omega) - G^{A}(q,\omega)) \tanh\frac{\beta\omega}{2}.
\end{eqnarray}

Full Green's functions are governed by the Dyson equation,
in which the tunneling matrix acts as an effective single-particle potential:
\begin{eqnarray}
\label{eq:GDyson}
  \Bigg[
  \begin{array}{cc}
    \check{G}_{LL} & \check{G}_{LR} \\
    \check{G}_{RL} & \check{G}_{RR}
  \end{array}
  \Bigg]
  =
  \Bigg[
  \begin{array}{cc}
    \check{G}_{LL}^{(0)} &  \\
     & \check{G}_{RR}^{(0)}
  \end{array}
  \Bigg]
  +
  \Bigg[
  \begin{array}{cc}
    \check{G}_{LL}^{(0)} &  \\
     & \check{G}_{RR}^{(0)}
  \end{array}
  \Bigg]
  \circ
  \Bigg[
  \begin{array}{cc}
     & \check{T} \\
    \check{T}^{\dagger} & 
  \end{array}
  \Bigg]
  \circ
  \Bigg[
  \begin{array}{cc}
    \check{G}_{LL} & \check{G}_{LR} \\
    \check{G}_{RL} & \check{G}_{RR}
  \end{array}
  \Bigg].
\end{eqnarray}
Here, $\check{G}_{LL}$, $\check{G}_{RR}$, and $\check{G}_{RL}$ are defined analogously to $\check{G}_{LR}$ in Eqs.~\eqref{eq:G11}-\eqref{eq:G22}.
The quantities $\check{G}^{(0)}_{LL}$ and $\check{G}^{(0)}_{RR}$ correspond to the Green's functions of $\mathcal{H}_L$ and $\mathcal{H}_R$, respectively.
In the rotated Keldysh space, the tunneling matrix is given by
\begin{eqnarray}
  \check{T} = \delta(t-t')\begin{pmatrix}
    T_{kq} & \\
    & T_{kq}
  \end{pmatrix}.
\end{eqnarray}
The circle product ``$\circ$" denotes integration over the internal variables:
\begin{eqnarray}
  [\check{G}\circ \check{T}\circ\check{G}](1,4) \equiv \sum_{k_2,k_3}\int dt_2dt_3~\check{G}(1,2) \check{T}(2,3) \check{G}(3,4).
\end{eqnarray}

By solving the Dyson equation for $\check{G}_{LR}$, we obtain
\begin{eqnarray}
  \check{G}_{LR} &=& \check{G}_{LL}^{(0)}\circ \check{T}^{\dagger}\circ \check{G}_{RR}^{(0)} + \check{G}_{LL}^{(0)}\circ \check{T}^{\dagger}\circ \check{G}_{RR}^{(0)}\circ \check{T} \circ \check{G}_{LR} \nonumber \\
  &=& \check{G}_{LL}^{(0)}\circ \check{\Sigma} \circ \check{G}_{RR}^{(0)},
  \label{eq:glr2}
\end{eqnarray}
where we have introduced the self-energy $\check{\Sigma}$, satisfying the following Dyson equation
\begin{eqnarray}
  \check{\Sigma} = \check{T}^{\dagger} + \check{T}^{\dagger}\circ \check{G}_{RR}^{(0)}\circ \check{T} \circ \check{G}_{LL}^{(0)} \circ \check{\Sigma}.
  \label{eq:sigma2}
\end{eqnarray}
Solving the Dyson equation for the self-energy yields the full Green's function across the junction.

\subsection{Fourier representations}
To solve the Dyson equations, we introduce Fourier representations.
In the right system, the Green's function depends only on the relative time $t-t'$, 
and its Fourier transform is given by
\begin{eqnarray}
  \check{G}_{RR}^{(0)}(t,t') = \int \frac{d\omega}{2\pi}~e^{-i\omega(t-t')}\check{G}^{(0)}_{RR}(\omega).
\end{eqnarray}
In contrast, in the superconducting case, Green's function in the left system explicitly depends on both $t$ and $t'$ due to the gauge potential induced by the voltage.
It takes the following form 
\begin{eqnarray}
  \check{G}^{(0)}_{LL}(t,t') = \begin{pmatrix}
    e^{-\frac{i}{\hbar}eV(t-t')}\check{g}_L(t-t') & e^{-\frac{i}{\hbar}eV(t+t')}\check{f}_L(t-t') \\
    e^{\frac{i}{\hbar}eV(t+t')}\check{f}_L^{\dagger}(t-t') & e^{\frac{i}{\hbar}eV(t-t')}\check{g}^{\dagger}_L(t-t') 
  \end{pmatrix},
\end{eqnarray}
where $\check{g}$ and $\check{g}^{\dagger}$ denote the normal Green's functions for particles and holes, respectively,
and $\check{f}$ and $\check{f}^{\dagger}$ represent the anomalous Green's functions, representing the Cooper pair amplitudes.
Note that in the normal state, where anomalous components vanish, 
the Green's function depends only on the relative time,
and thus a standard Fourier transform applies.
To handle the time dependence in the superconducting case of the left system,
we introduce the discrete frequency component $\Omega_n = 2eVn/\hbar$,
and perform the mixed Fourier transform as
\begin{eqnarray}
  &&\check{G}^{(0)}_{LL}(t,t') = \int \frac{d\omega}{2\pi}~\sum_n e^{-i\omega(t-t')}e^{-i\Omega_n t}\check{G}^{(0)}_{LL}(\omega,\Omega_n), \\
  &&\check{G}^{(0)}_{LL}(\omega,\Omega_n) = \begin{pmatrix}
    \delta_{n,0}~\check{g}_L(\omega-eV/\hbar) & \delta_{n,1}~\check{f}_L(\omega+eV/\hbar) \\
    \delta_{n,-1}~\check{f}_L^{\dagger}(\omega-eV/\hbar) & \delta_{n,0}~\check{g}^{\dagger}_{L}(\omega+eV/\hbar)
  \end{pmatrix},
\end{eqnarray}
where $\delta_{i,j}$ is the Kronecker delta.
Applying the same transformation to the self-energy, 
the Dyson equation in Eq.~\eqref{eq:sigma2} becomes
\begin{eqnarray}
  \check{\Sigma}(\omega,\Omega_n) = \delta_{n,0}\check{T}^{\dagger} + \check{T}^{\dagger}\circ \check{G}^{(0)}_{RR}(\omega+\Omega_n)\circ\check{T}\circ\sum_{n'}\check{G}^{(0)}_{LL}(\omega+\Omega_n',\Omega_n-\Omega_n')\circ\check{\Sigma}(\omega,\Omega_n'),
  \label{eq:sigma}
\end{eqnarray}
where the circle product implies summation over the momentum variables.
Substituting this into the expression for the Green's function at equal time in Eq.~\eqref{eq:glr2}, we obtain
\begin{eqnarray}
  \check{G}_{LR}(t,t) = \sum_{n} e^{-i\Omega_n t}\int \frac{d\omega}{2\pi}\sum_{n'}\check{G}_{LL}^{(0)}(\omega+\Omega_n',\Omega_n-\Omega_n')\circ\check{\Sigma}(\omega,\Omega_n')\circ \check{G}_{RR}^{(0)}(\omega).
\end{eqnarray}
The steady tunneling current arises from the $\Omega_n = 0$ components.

When electron tunneling occurs locally in real space, as in STM experiments,
the tunneling matrix can be treated as momentum-independent,
which simplifies the expression for the tunneling current:
\begin{eqnarray}
  \label{eq:Ipoint}
  I(V,t) = -2e|T| \Re ~\text{tr}~\langle G^<_{LR}(t,t)\rangle_{k,q},
\end{eqnarray}
where the momentum-averaged Green's function is defined as $\langle G^<_{LR}(t,t') \rangle_{k,q} = \sum_{k,q}G_{LR}^<(k,t;q,t')$.
The Dyson equation governing the momentum-averaged Green's function then takes the form
\begin{eqnarray}
  \langle \check{G}^{LR}(t,t)\rangle_{k,q} = \sum_{n} e^{-i\Omega_n t}\int \frac{d\omega}{2\pi}~\sum_{n'} 
  \langle \check{G}_{LL}^{(0)}(\omega+\Omega_n',\Omega_n-\Omega_n')\rangle_k\langle {\check{\Sigma}}(\omega,\Omega'_{{n}})\rangle_{k,q}\langle \check{G}^{(0)}_{RR}(\omega) \rangle_q,
\end{eqnarray}
where the self-energy satisfies
\begin{eqnarray}
  \langle\check{\Sigma}(\omega,\Omega_n)\rangle_{k,q} = |T|\delta_{n,0} + |T|^2\langle\check{G}^{(0)}_{RR}(\omega+\Omega_n)\rangle_q\sum_{n'}\langle\check{G}^{(0)}_{LL}(\omega+\Omega_n',\Omega_n-\Omega_n')\rangle_k\langle{\check{\Sigma}}(\omega,\Omega_n')\rangle_{k,q}.
\end{eqnarray}
Here, we define the momentum-averaged bare Green's functions as $\langle \check{G}^{(0)}_{LL}(\omega,\Omega_n) \rangle_{k} = \sum_{k}\check{G}^{(0)}_{LL}(k,\omega,\Omega_n)$ and
$\langle \check{G}^{(0)}_{RR}({\omega}) \rangle_{q} = \sum_{q}\check{G}^{(0)}_{{RR}}(q,\omega)$.

\subsection{Andreev current in STM probing topological superconductors with $s$-wave tips}
\begin{figure}[t]
  \centering
  \includegraphics[width=\linewidth]{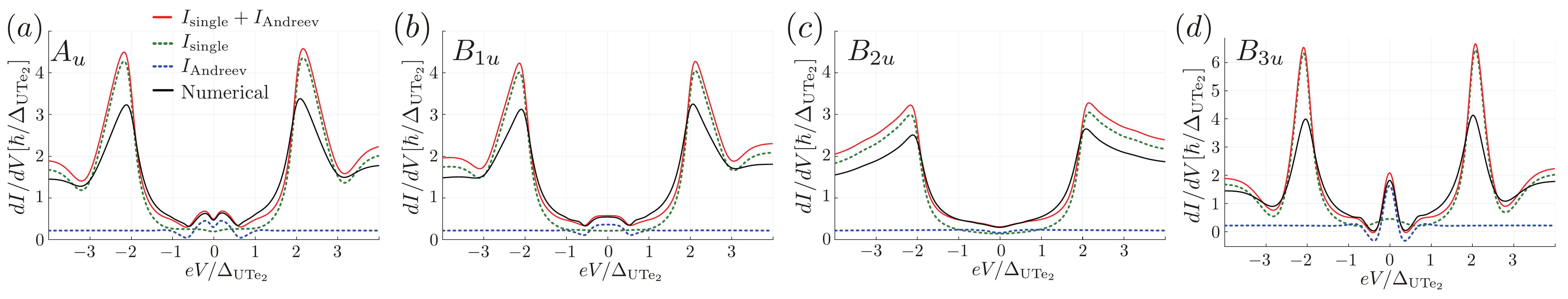}
  \caption{Comparison between numerical calculations and analytical expressions.
  The tunneling amplitude is set to $\alpha=0.2$.
  The black solid curves represent numerically calculated $dI/dV$ spectra, incorporating tunneling processes to all orders in perturbation theory.
  The dotted green curves correspond to the analytical result of single-particle tunneling [Eq.~\eqref{eq:Isingle}].
  The dotted blue curves correspond to the analytical result of Andreev tunneling current [Eq.~\eqref{eq:IAndreev}].
  The red curves correspond to the sum of the contributions from Eq.~\eqref{eq:Isingle} and Eq.~\eqref{eq:IAndreev}.
  In the low-bias regime, the $dI/dV$ characteristics are governed by the Andreev tunneling current.
  In contrast, in the high-bias regime $(eV>\Delta_s)$, the single-particle tunneling contribution becomes dominant. 
  }
  \label{fig:dIdV_ana}
\end{figure}
We consider a point-contact junction in which a topological superconductor is placed on the right side and an $s$-wave superconductor on the left side.
For notation simplicity,
we denote the momentum-averaged Green's function as $G(\omega)$ throughout the following discussion.
Expanding Eq.~\eqref{eq:Ipoint} to the second order in the tunneling matrix element $T$,
the tunneling current is given by 
\begin{eqnarray}
\label{eq:I2}
  I^{(2)}(V,t) = -2|T|^2e \Re~\text{tr} \int \frac{d\omega}{2\pi}~\Big\{[\check{g}_L(\omega-eV/\hbar)\check{g}_R(\omega)]^< - e^{-2ieVt/\hbar}[\check{f}_L(\omega+eV/\hbar)\check{f}_R^{\dagger}]^<\Big\}.
\end{eqnarray}
The first term represents the contribution of the single particle tunneling $I_{\rm single}$.
By straightforward calculation, we arrive at a familiar form
\begin{eqnarray}
\label{eq:Isingle}
    I_{\rm single} = (4\pi T^2e/\hbar) \int_0^{eV}dE~\rho_{\rm tip}(E-eV)\rho_{\rm sample}(\omega),
\end{eqnarray}
where $\rho_{\rm tip}$ and $\rho_{\rm sample}$ correspond to the DOS for the left and right systems, respectively.
Here, we have used spectrum representation of the retarded/advanced Green's function
\begin{eqnarray}
  &&g^{R/A}(\omega) = \int dE~\frac{\rho(E)}{\hbar\omega\pm i\delta - E}, \\
  &&g^{\dagger R/A}(\omega) = \int dE~\frac{\rho(E)}{\hbar\omega\pm i\delta + E}.
\end{eqnarray}
However, since an $s$-wave superconducting tip has no quasiparticle states within the superconducting gap,
this contribution vanishes.
The second term in Eq.~\eqref{eq:I2} corresponds to the Josephson current.
In this case, since the junction is formed between a spin-triplet and a spin-singlet superconductor, the orthogonality in spin space leads to the vanishing of this term as well. 
However, if SOC coupling or magnetic impurities are present, a finite Josephson current may be induced,
although it is generally expected to be small.

The leading contribution to the steady tunneling current in a junction between a conventional superconductor and a topological superconductor therefore
arises from fourth-order tunneling processes.
The dominant contribution is the Andreev tunneling current, which originates from Andreev reflection and is given by
\begin{eqnarray}
  I_{\rm Andreev} = -4e|T|^4\Re \int \frac{d\omega}{2\pi}~[\check{f}_L(\omega)\check{g}_R^{\dagger}(\omega-eV/\hbar)\check{f}_{L}^{\dagger}(\omega)\check{g}_R(\omega+eV/\hbar)]^<.
\end{eqnarray}
In this process, electrons and holes in the surface state of the topological superconductor are converted into Cooper pairs in the $s$-wave superconductor and vice versa through tunneling.
We employ the analytical expression for the anomalous Green's function of the $s$-wave superconductor:
\begin{eqnarray}
  f_L^{R/A}(\omega) = f_L^{\dagger R/A}(\omega) = -\frac{N_L\pi\Delta_s}{\sqrt{\Delta^2-(\hbar\omega\pm i\delta)^2}},
\end{eqnarray}
where $N_L$ denotes the normal-state DOS at the Fermi level, and $\Delta_s$ is the superconducting gap amplitude.
By straightforward calculation, we arrive at the final expression:
\begin{eqnarray}
\label{eq:IAndreev}
  I_{\rm Andreev}(V) = (4e\pi^3|T|^4N_L^2/\hbar)\int_0^{2eV}dE~\rho_{\rm sample}(E-2eV)\rho_{\rm sample}(E).
\end{eqnarray}
\end{widetext}
It is noteworthy that the Andreev current is independent of the superconducting gap $\Delta_s$.

Figure~\eqref{fig:dIdV_ana} compares the numerical calculations with the analytical expressions.
The black solid curves represent the numerically computed $dI/dV$ spectra, 
incorporating tunneling processes to all orders in perturbation theory.
The red curves correspond to the analytical results, which are the sum of contributions from 
single-particle tunneling current [Eq.~\eqref{eq:Isingle}], shown by green dotted lines,
and Andreev tunneling current [Eq.~\eqref{eq:IAndreev}], shown by blue dotted lines.
Note that, due to the presence of a small but finite smearing factor, 
single-particle tunneling contributes even within the gap of the $s$-wave superconductor.
In the low-bias regime, the analytical expressions agree well with the numerical results.
Moreover, the peak structure at low bias is predominantly governed by Andreev tunneling current.
When the DOS is large and the tunneling probability is high, 
higher-order processes can act as self-energy corrections that suppress the tunneling current.
As a result, the coherence peaks of the $s$-wave superconductor, which appear at $eV = 2\Delta_{\text{UTe}_2}$, are more suppressed in the numerical results than in the analytical results, which only account for contributions up to the fourth order.
Similarly, the ZBP for the $B_{3u}$ state [Fig.~\ref{fig:dIdV_ana}(d)] also exhibits suppression in the numerical calculation.

\section{Two-orbital model\label{App:B}}
\begin{figure}[t]
  \centering
  \includegraphics[width=\linewidth]{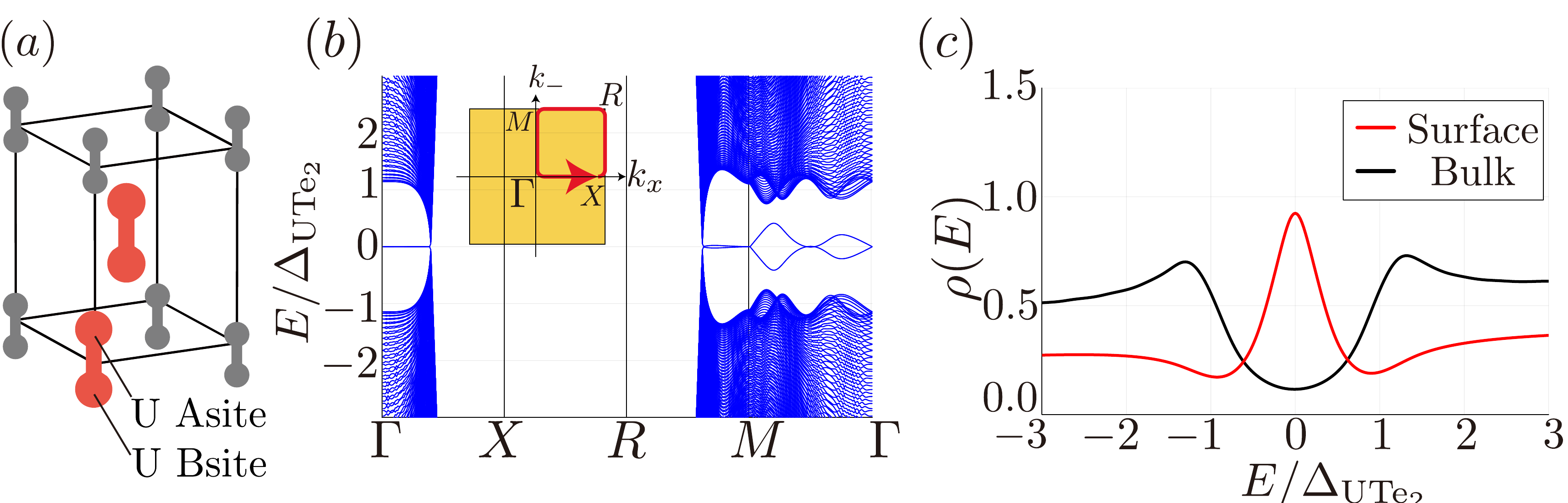}
  \caption{(a)~Crystal structure of UTe$_2$. Only uranium sites are shown.
  (c)~Quasiparticle energy bands for the $B_{3u}$ state in the two-orbital model.
     The staggered Rashba spin-orbit coupling, originating from local inversion symmetry breaking at uranium sites, has little effect on the zero-energy states.
  (d)~Surface DOS exhibiting a pronounced zero-energy peak due to the presence of the 2D nearly 2D flat band.
  }
  \label{fig:B3u_2}
\end{figure}
In this section, we demonstrate that the 2D ABS persists for the $B_{3u}$ state even when employing a more realistic model that incorporates key features of UTe$_2$, beyond the simplified model employed in the main text.
Figure~\ref{fig:B3u_2}(a) shows the crystal structure of UTe$_2$.
UTe$_2$ crystallizes in a body-centered orthorhombic structure with $D_{2h}$ point group symmetry.
Within each unit cell, uranium atoms are aligned along the crystal $c$-axis in a dimer-like configuration, resulting in local inversion symmetry breaking at the uranium sites.

We adopt a two-orbital tight-binding model that includes the site degree of freedom associated with uranium atoms~\cite{tei2023possible,shishidou2021topological}.
The normal-state Hamiltonian, which respects the body-centered orthorhombic symmetry, is given by
\begin{eqnarray}
    H_{\rm N}(\bm{k}) = \epsilon_0(\bm{k}) - \mu + f_x(\bm{k})\rho_x + f_y(\bm{k})\rho_y + \bm{g}(\bm{k})\cdot\boldsymbol{\sigma}\rho_z , \nn \\
\end{eqnarray}
where
\begin{align}
    &\epsilon_0(\bm{k}) = 2t_1 \cos k_x + 2t_2 \cos k_y, \\
    &f_x(\bm{k}) = t_3 + t_4\cos(k_x/2)\cos(k_y/2)\cos(k_z/2), \\
    &f_y(\bm{k}) = t_5\cos(k_x/2)\cos(k_y/2)\sin(k_z/2), \\
    &g_x({\bm k}) = R_x \sin k_y, \\
    &g_y({\bm k}) = R_y \sin k_x,\\
    &g_z({\bm k}) = R_z\sin(k_x/2)\sin(k_y/2)\sin(k_z/2).
\end{align}
Here, $\boldsymbol{\sigma}$ and $\boldsymbol{\rho}$ are the Pauli matrices acting on spin and uranium-site spaces, respectively.
The last term represents Rashba-type spin-orbit coupling (SOC) arising from local inversion symmetry breaking at the uranium site.
To reproduce a cylindrical Fermi surface, we use the following parameters:
$\mu = -1.8$, $t_1=-0.5$, $t_2 = 0.375$, $t_3=-0.7$, $t_4 = 0.65$, $t_5 = -0.65$, $R_x = 0.1$, $R_y = 0.1$, $R_z = 0.1$.

The gap function also incorporates the site degree of freedom.
We consider an inter-site spin-triplet Cooper pair, described by
\begin{eqnarray}
    \Delta(\bm{k}) = \begin{pmatrix}
        0 & \bm{d}({\bm k})\cdot\boldsymbol{\sigma}i\sigma_y \\
        \bm{d}({\bm k})\cdot\boldsymbol{\sigma}i\sigma_y & 0
    \end{pmatrix},
\end{eqnarray}
where the matrix structure refers to the uranium-site space.
The $d$-vector for the $B_{3u}$ state is given by $\bm{d} = \begin{pmatrix}
    0, & C_2 \sin k_y, & C_3\sin k_z
\end{pmatrix}$,
as also used in the main text.
We set $C_2 = 0.05$ and $C_3 = 0.05$ in the calculations.

Figure~\ref{fig:B3u_2}(b) shows the quasiparticle energy bands for the $B_{3u}$ state in the two-uranium-site model,
obtained by diagonalizing a slab system with open $(011)$ surfaces.
Along the $M-\Gamma$ line, in-gap states appear as a result of nontrivial Berry phases.
Furthermore, even in the presence of staggered Rashba SOC presence,
zero-energy states originating from weak spin conservation appear along the $\Gamma-X$ and $R-M$ lines.
As a result, the surface exhibits a large zero-energy DOS, as shown in Fig.~\ref{fig:B3u_2}(c).

\bibliography{paper}

\end{document}